\documentclass[a4paper,10pt]{article}
\pdfoutput=1 

\usepackage{jheppub} 

\usepackage[T1]{fontenc} 

\usepackage{setspace}

\usepackage{subcaption}
\usepackage{physics}
\usepackage{dsfont}
\usepackage{tensor}
\usepackage[normalem]{ulem}
\usepackage{xcolor}
\usepackage{graphicx}
\usepackage[export]{adjustbox}
\usepackage{BOONDOX-cal}

\graphicspath{{../pics/}}
\usepackage{tikz}
\usetikzlibrary{quantikz}
\usepackage{yquant}

\colorlet{darkblue}{blue!70!black}
\colorlet{darkgreen}{green!50!black}

\newcommand{\spec}{\operatorname{spec}}

\title{\boldmath A Type $I$ Approximation of the Crossed Product}

\author{Ronak M Soni}
   \affiliation{Department of Applied  Mathematics and Theoretical Physics, University of Cambridge,\\Wilberforce Road, Cambridge, CB3 0WA, United Kingdom}
\emailAdd{rs2194@damtp.cam.ac.uk}

\abstract{
  I show that an analog of the crossed product construction that takes type $III_{1}$ algebras to type $II$ algebras exists also in the type $I$ case.
  This is particularly natural when the local algebra is a non-trivial direct sum of type $I$ factors.
  Concretely, I rewrite the usual type $I$ trace in a different way and renormalise it.
  This new renormalised trace stays well-defined even when each factor is taken to be type $III$.
  I am able to recover both type $II_{\infty}$ as well as type $II_{1}$ algebras by imposing different constraints on the central operator in the code.
  An example of this structure appears in holographic quantum error-correcting codes; the central operator is then the area operator.
}

\begin{document}
\maketitle
\flushbottom

\section{Introduction} \label{sec:intro}
Recently, the old observation \cite{Susskind:1994sm,Solodukhin:2011gn,Bousso:2015mna} that generalised entropy --- the sum of gravitational entropy and the entropy of bulk matter fields --- is better defined than each term in the sum has been placed on firmer footing \cite{Witten:2021unn,Bahiru:2022mwh,Chandrasekaran:2022eqq,Chandrasekaran:2022cip,Penington:2023dql,Kolchmeyer:2023gwa,AliAhmad:2023etg,Jensen:2023yxy,Klinger:2023tgi,Ali:2023bmm} using the crossed product construction \cite{turumaru1958crossed}.
I will refer to this construction in the context of gravity as the `gravitational crossed product,' even though it is mathematically not distinct, as I would like to make physical comments that pertain only to this case.
See \cite{Sorce:2023fdx} for a recent introduction aimed at high energy theorists.

The gravitational crossed product takes the type $III_{1}$ algebra of bulk matter fields \cite{Leutheusser:2021qhd,Leutheusser:2021frk,Leutheusser:2022bgi} and adjoins a `boundary' Hamiltonian.
In the asymptotically AdS case this boundary Hamiltonian is the ADM Hamiltonian and in the asymptotically dS/arbitrary subregion of the bulk case it is the (modular) Hamiltonian of an observer; what is important is the existence of a state in which the bulk modular Hamiltonian is related to this boundary Hamiltonian by a constraint.
The resulting algebra is type $II$.
Unlike in type $III$ algebras, type $II$ algebras have well-defined entropies, and it can be shown that differences of this type $II$ entropy between different states is the difference of the generalised entropies between those states \cite{Chandrasekaran:2022eqq,Ali:2023bmm}.
Naively, this construction seems to rely crucially on the existence of an outer automorphism of the local algebra of bulk matter fields; these do not exist for type $I$ factors and so it seems that this approach is `purely infinite.' 

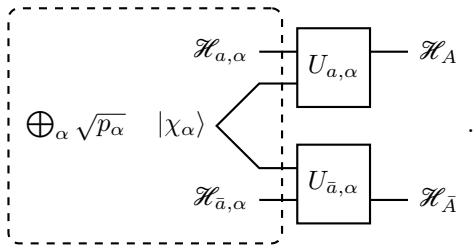
\begin{figure}[h!]
  \centering
    \begin{quantikz}[align equals at=2.5]
    \gategroup[4,steps=7,style={dashed,rounded corners,inner xsep=5pt}]{}
    & & & & & &\lstick{$\mathcal{H}_{a,\alpha}$} & \gate[wires=2]{U_{a,\alpha}} & \qw \rstick{$\mathcal{H}_{A}$} \\[-.7cm]
    & & & & & &\makeebit{$\bigoplus_{\alpha} \sqrt{p_{\alpha}} \quad \ket{\chi_{\alpha}}$} & & \\
    & & & & & & & \gate[wires=2]{U_{\bar{a},\alpha}} & \\ [-.7cm]
    & & & & & & \lstick{$\mathcal{H}_{\bar{a},\alpha}$} & & \qw \rstick{$\mathcal{H}_{\bar{A}}$}
  \end{quantikz}.
  \caption{A holographic QEC, as introduced in \cite{Harlow:2016vwg}. The boxed part is just a system such that the local algebra is a type $I$ non-factor algebra.}
  \label{fig:hol-qec}
\end{figure}

A similar argument has been made in the context of Harlow's holographic quantum error-correcting (QEC) code \cite{Harlow:2016vwg},\footnote{See \cite{Faulkner:2022ada} for a more continuum-like treatment of holographic QEC.} illustrated in figure \ref{fig:hol-qec}, by \cite{Gesteau:2023hbq}.
\cite{Gesteau:2023hbq} shows the existence of a conditional expectation in these codes that keeps entanglement entropy constant while moving some of it from the bulk to the area operator.
Perhaps surprisingly, \cite{Gesteau:2023hbq} only uses type $I$ algebras to make this argument.

This work is the bridge between the two arguments.
I argue that a construction that contains the `physical' aspects of the crossed product construction exists even in type $I$ algebras.
While this is true in any such situation, this construction is particularly natural when the local algebra is a non-trivial direct sum of type $I$ factors.
An example of this structure is in a holographic QEC.
The results of \cite{Klinger:2023tgi}, though different in detail and perspective, have the same flavour: they show that the algebras of the quantisation of classical systems with edge modes have a crossed-product structure.

A specific reason this is an interesting observation involves tensor network toy models for holographic states.
In these toy models, the bulk algebra is always type $I$; the observation in this paper shows how to define an entanglement entropy in these models that is free of bulk UV divergences.
We will expand on this in upcoming work \cite{Akers:2023}.
  Another reason this is interesting is in understanding the connection between the path integral proof of the well-defined-ness of the generalised entropy and the algebraic proof.

The plan of the paper is the following.
I begin with a trivial observation in section \ref{sec:idea}.
I move on to a small discussion of the brick wall boundary condition that is commonly used to approximate entanglement entropy in the type $III$ algebras corresponding to subregions of QFTs in section \ref{sec:type-1}; this section can be safely skipped by most readers.
In section \ref{sec:main-construction} I describe the central result.
I end with discussion in section \ref{sec:conc}.

\section{The Basic Idea} \label{sec:idea}
There are two basic ideas that go into the main construction of this work.
The first concerns the most important difference between the holographic QEC model and the crossed-product construction.
And the second is a useful reformulation of the trace in any algebra where it exists.
Neither is new, but together they suggest the construction of section \ref{sec:main-construction}.

\paragraph{Area and `Boundary' Energy}
The holographic QEC and the gravitational crossed product share the following structure.
They both involve embedding `bulk' algebras on the left and the right into a larger algebra involving an extra degree of freedom.
The differences are as follows.
Firstly, the bulk algebra is type $I$ in the holographic QEC whereas it is type $III$ in the gravitational crossed product; this will be the subject of section \ref{sec:main-construction}.
The more physical difference is the extra degree of freedom: in the holographic QEC it is the area, which is a central operator supported on the co-dimension-two HRT surface, while in the gravitational crossed product it is the ADM energy (in the asymptotically AdS case when the boundary subregion is an entire CFT).
More generally, the ADM energy is replaced by the modular energy of some boundary observer \cite{Chandrasekaran:2022cip,Jensen:2023yxy,AliAhmad:2023etg}; I will refer to it as the `boundary' energy.

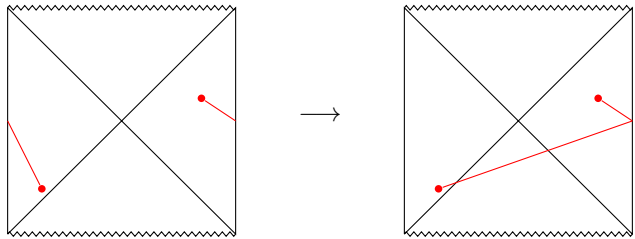
\begin{figure}[h!]
  \centering
  \begin{tikzpicture}[scale=1.5,baseline={(current bounding box.center)}]
    \draw (0,0) -- (0,2) -- (2,0) -- (2,2) -- (0,0);
    \draw[decorate,decoration={zigzag,amplitude=.3mm,segment length=1mm}] (0,2) -- (2,2);
    \draw[decorate,decoration={zigzag,amplitude=.3mm,segment length=1mm}] (0,0) -- (2,0);
    \node[circle,fill,inner sep=1,red] (r) at (1.7,1.2) {};
    \node[circle,fill,inner sep=1,red] (l) at (.3,.4) {};
    \draw[red] (r) -- (2,1);
    \draw[red] (l) -- (0,1);
  \end{tikzpicture}
  \qquad $\longrightarrow$ \qquad
  \begin{tikzpicture}[scale=1.5,baseline={(current bounding box.center)}]
    \draw (0,0) -- (0,2) -- (2,0) -- (2,2) -- (0,0);
    \draw[decorate,decoration={zigzag,amplitude=.3mm,segment length=1mm}] (0,2) -- (2,2);
    \draw[decorate,decoration={zigzag,amplitude=.3mm,segment length=1mm}] (0,0) -- (2,0);
    \node[circle,fill,inner sep=1,red] (r) at (1.7,1.2) {};
    \node[circle,fill,inner sep=1,red] (l) at (.3,.4) {};
    \draw[red] (r) -- (2,1);
    \draw[red] (l) -- (2,1);
  \end{tikzpicture}
  \caption{Changing the extra degree of freedom is analogous to changing the dressing. On the left is the dressing relevant for the holographic QEC; nothing is dressed across the HRT surface and so one-sided operators cannot change the area. On the right is that relevant for the gravitational crossed product; everything is dressed to the right boundary and so the left ADM energy is central. The trace for the left algebra is best written down by changing the dressing to be to the left boundary. In other words, the state is dressed as in the right picture but the operators are dressed as in the left picture.}
  \label{fig:re-dressing}
\end{figure}

These two quantities are related by the modular energy of the bulk fields in the entanglement wedge, which is known in the literature as the JLMS relation \cite{Wald:1993nt,Jafferis:2015del}.
So the first step in the construction will be relabelling a basis of the holographic QEC.
The labels for the basis elements need to be changed from the pair (area,bulk modular energy) to (bulk modular energy, `boundary' energy).
This is a little subtle in the general case, involving a set of simultaneous non-linear equations \eqref{eqn:X-defn}, but becomes as simple as it sounds when I mock up the gravitational case, in \eqref{eqn:X-defn-simple}.
It would be interesting to restate this as a change in quantum reference frame, see e.g. \cite{Aharonov:1967zza,Vanrietvelde:2018pgb}.
More colloquially, it can also be thought of as a change in dressing \cite{Chandrasekaran:2022eqq}, as shown in figure \ref{fig:re-dressing}.

More precisely, I find that the dressing is changed as in figure \ref{fig:re-dressing} in the sense that the different sectors are indexed by a fixed value of the left `boundary'  Hamiltonian, but that the same is not true for the algebra.
Notice that in the re-dressing depicted, only the left-sided operators change.
The crossed product algebra for right-sided operators are dressed to the right as shown, but that for left-sided operators are dressed to the \emph{left} \cite{Chandrasekaran:2022eqq}.
In other words, as a far as the algebras are concerned, the dressing is as on the left of the figure.
The only thing that changes for the algebras is that the sectors for the right (left) algebra are labelled by the values of the right (left) `boundary' Hamiltonian instead of the area.
This provides moral evidence for the idea \cite{Klinger:2023tgi} that the crossed product construction is essentially the type $III$ version of an algebra with a centre.

Allowing the local algebra to have a centre and changing dressing as above is the most important step in approximating the crossed product.\footnote{Even when we begin without a centre, the change in dressing requires adding it in, as we will see.}
The crossed product construction takes a type $III_{1}$ algebra and, using the fact that the modular automorphism is outer, constructs a type $II$ algebra with the trace \eqref{eqn:type-2-tr}.
While both of these algebras, the initial and the crossed product, are hyper-finite, the procedure that takes the initial algebra to the crossed product is not trivial to approximate.
This is because type $I$ factors have no outer automorphisms, and the crossed product construction is trivial when built on an inner automorphism.
However, as it turns out, the physical aspects don't depend on this mathematical detail.

\paragraph{Traces}
The second observation regards the trace.
Consider a state $\ket{\psi} \in \mathcal{H}_{\mathrm{L}} \otimes \mathcal{H}_{\mathrm{R}}$ that is cyclic and separating with respect to $B\left( \mathcal{H}_{\mathrm{L},\mathrm{R}} \right)$.
This essentially means that $\mathrm{dim}\, \mathcal{H}_{\mathrm{L}} = \mathrm{dim}\, \mathcal{H}_{\mathrm{R}}$ and the Schmidt spectrum of the state has no zeros.
Then, the trace of an operator $O_{\mathrm{R}} \in B\left( \mathcal{H}_{\mathrm{R}} \right)$ can be rewritten as
\begin{align}
  \tr_{\mathcal{H}_{\mathrm{R}}} O_{\mathrm{R}} &= \tr_{\mathcal{H}_{\mathrm{R}}} \left( \rho_{\mathrm{R}} O_{\mathrm{R}} \rho_{\mathrm{R}}^{-1} \right) \nonumber\\
  &= \mel{\psi}{O_{\mathrm{R}} \rho_{\mathrm{R}}^{-1}}{\psi}.
  \label{eqn:tr-triv}
\end{align}

This is entirely analogous to the trace in the crossed-product algebra.
In that case, there are two type $II$ algebras $\mathcal{A}_{\mathrm{R}}, \mathcal{A}_{\mathrm{R}}' = \mathcal{A}_{\mathrm{L}}$, the modular operator is written as
\begin{equation}
  \Delta_{\psi} = K \tilde{K}^{-1}, \qquad K \in \mathcal{A}_{\mathrm{R}}, \tilde{K} \in \mathcal{A}_{\mathrm{L}}
  \label{eqn:type-2-mod}
\end{equation}
and the trace is
\begin{equation}
  \tr O_{\mathrm{R}} = \mel{\psi}{O_{\mathrm{R}} K^{-1}}{\psi}.
  \label{eqn:type-2-tr}
\end{equation}

Comparing the modular operator for type $I$ factors, $\Delta_{\psi} = \rho_{\mathrm{R}} \rho_{\mathrm{L}}^{-1}$, with \eqref{eqn:type-2-mod}, one may intuit that $K,\tilde{K}$ are the reduced density matrices for $\mathcal{A}_{\mathrm{R}}, \mathcal{A}_{\mathrm{L}}$.
This is in fact true.
Consider
\begin{align}
  \tr K O_{\mathrm{R}} &= \mel{\psi}{ K O_{\mathrm{R}} K^{-1}}{\psi} \nonumber\\
  &= \mel{\psi}{ K \tilde{K}^{-1} O_{\mathrm{R}} \tilde{K} K^{-1}}{\psi} \nonumber\\
  &= \mel{\psi}{O_{\mathrm{R}}}{\psi}.
  \label{eqn:K-rho}
\end{align}
In the second equality, I have used the fact that $\tilde{K} \in \mathcal{A}_{\mathrm{L}}$ commutes with $O_{\mathrm{R}} \in \mathcal{A}_{\mathrm{R}}$, and in the third I have used the fact that $\Delta_{\psi} \ket{\psi} = \ket{\psi}$.
So $K$ reproduces the correct expectation values for all operators in $\mathcal{A}_{\mathrm{R}}$, meaning that it is the reduced density matrix.

\section{Type $I$ Approximations and Type $III$ Limits} \label{sec:type-1}
In the main section of this paper, I will imagine the same physical subsystem of the bulk matter QFT to have both (approximate) type $I$ algebras and exact type $III$ algebras.
While this is standard in the literature on entanglement in QFT, it is also somewhat implicit.
So, I include a small introduction here for non-experts.

\subsection{An Example of a Type $I$ Approximation} \label{ssec:eg}
Consider the example of a free massless scalar field $\phi(x)$ on a two-dimensional asymptotically AdS two-sided eternal black hole background, with metric (in one exterior)
\begin{equation}
  ds^{2} = - \frac{r^{2} - r_{h}^{2}}{\ell^{2}} dt^{2} + \frac{\ell^{2} dr^{2}}{r^{2} - r_{h}^{2}}.
  \label{eqn:2d-bh-metric}
\end{equation}
In this section, this geometry is merely a geometry on which the QFT lives; there is no gravity in  the theory.
Impose a brick wall boundary condition at a stretched horizon $r - r_{h} = \delta \ll \ell$,
\begin{equation}
  \phi (r = r_{h} + \delta) = \phi_{0},
  \label{eqn:brick-wall}
\end{equation}
and the usual normalisable boundary condition at the asymptotic boundary,
\begin{equation}
  \phi \xrightarrow{r \to \infty} \phi_{1}.
  \label{eqn:asymp-bd}
\end{equation}

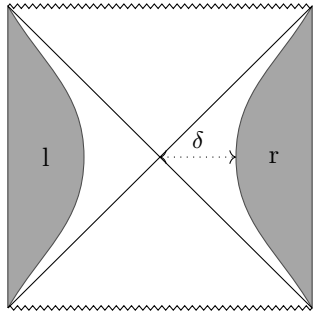
\begin{figure}[h!]
  \centering
  \begin{tikzpicture}[scale=2]
	  \draw (0,0) -- (0,2) -- (2,0) -- (2,2) -- (0,0);
	  \draw[decorate,decoration={zigzag,amplitude=.3mm,segment length=1mm}] (0,2) -- (2,2);
	  \draw[decorate,decoration={zigzag,amplitude=.3mm,segment length=1mm}] (0,0) -- (2,0);
	  \draw[fill=gray,opacity=.7] (2,0) to[out=120,in=-90] (1.5,1) to[out=90,in=-120] (2,2);
	  \draw[fill=gray,opacity=.7] (0,0) to[out=60,in=-90] (.5,1) to[out=90,in=-60] (0,2);
	  \draw[<->,dotted] (1,1) to node[midway,above] {\small $\delta$} (1.5,1);
	  \node at (1.75,1) {$\mathrm{r}$};
	  \node at (0.25,1) {$\mathrm{l}$};
	\end{tikzpicture}
  \caption{I consider a scalar field on a two-sided black hole background with a brick wall.}
  \label{fig:brick-wall}
\end{figure}

Now, mode expand the field in terms of the solutions of the equation of motion.
Defining the alternate radial coordinate $\rho = \int_{\infty}^{r} \tfrac{\ell^{2}}{r^{2} - r_{h}^{2}} dr = \tfrac{\ell}{r_{h}} \coth^{-1} \tfrac{r}{r_{h}}$, the equation of motion is simply $\left( \partial_{t}^{2} - \partial_{\rho}^{2} \right) \phi = 0$.
In this co-ordinate, the stretched horizon is at $\rho = \rho_{s} \equiv \tfrac{\ell}{2 r_{h}} \log \tfrac{r_{h}}{\delta}$ and the asymptotic boundary is at $\rho = 0$.
The mode expansion of the field in the right exterior is
\begin{equation}
  \phi = \phi_{1} + \frac{\phi_{0} - \phi_{1}}{\rho_{s}} \rho + \sum_{n \in \mathds{N}} \left[ a_{\mathrm{r},\omega_{n}} e^{- i \omega_{n} t} + a_{\mathrm{r},\omega_{n}}^{\dagger} e^{i \omega_{n} t} \right] N_{\omega_{n}} \sin \left( \omega_{n} \rho \right), \quad \omega_{n} \equiv  \frac{\pi n}{\rho_{s}} = \frac{2 \pi n}{\frac{\ell}{r_{h}} \log \frac{r_{h}}{\delta}}.
  \label{eqn:mode-exp}
\end{equation}
Here, $N_{\omega_{n}}$ is a normalisation factor that will not be important for us.
The mode expansion on the left is the same except $a_{\mathrm{r},\omega_{n}} \to a_{\mathrm{l},\omega_{n}}$.
The state in the exterior of the two stretched horizons is the thermofield double state with inverse temperature $\beta = \tfrac{2\pi \ell^{2}}{r_{h}}$.
\begin{equation}
  \ket{\beta}_{\delta} = \bigotimes_{\omega_{n}} \sqrt{1 - e^{- \beta \omega_{n}}} e^{- \frac{\beta}{2} \omega_{n} a_{\mathrm{l},\omega_{n}}^{\dagger} a_{\mathrm{r},\omega_{n}}^{\dagger} } \ket{0}_{\mathrm{l},\omega_{n}} \ket{0}_{\mathrm{r},\omega_{n}}.
  \label{eqn:reg-hh}
\end{equation}
In the limit $\delta \to 0$ this becomes the boost-invariant Hartle-Hawking state.

The important outcome of the mode expansion \eqref{eqn:mode-exp} is that for any $\delta \neq 0$, one can assign a separable Hilbert space to the right exterior, that of one oscillator for the countably infinite set of frequencies $\omega_{n}$; this is similar to the Hilbert space of a free QFT on a circle.
The right algebra $\mathcal{a}_{\mathrm{r}}^{(\delta)}$ is the algebra of this countably infinite set of oscillators and is therefore type $I$; further, since it is generated by non-commuting pairs $a_{\mathrm{r},\omega_{n}}, a_{\mathrm{r},\omega_{n}}^{\dagger}$, it has no centre and it is a type $I$ factor.
One can also sum over different values of the Dirichlet boundary condition \eqref{eqn:brick-wall} at the stretched horizon; then the algebra is one copy of the above type $I$ factor for every value of $\phi_{0}$ and there is a centre generated by the operator that measures $\phi_{0}$ \cite{Lin:2018bud,Hung:2019bnq}.

Of course, the algebra of interest is actually the one at $\delta = 0$.\footnote{And with no constraint on $\phi(\delta)$ --- this is why one in general wants to sum over all values of $\phi_{0}$.}
However, the algebra of the exterior is not continuous in this limit.
To see this, consider taking the $\delta \to 0$ limit of the mode expansion \eqref{eqn:mode-exp}.
For simplicity, I take $t = 0$ and $\phi_{0} = \phi_{1} = 0$.
When $\delta \to 0$, the spacing between frequencies vanishes and the set of frequencies becomes the positive real line $\mathds{R}^{+}$; as a result, the sum needs to be converted to an integral,
\begin{align}
  \phi (\rho,t=0) &= \sum_{\omega_{n}} \Delta \omega \left[ \frac{a_{\mathrm{r},\omega_{n}}}{\Delta \omega} + \frac{a_{\mathrm{r},\omega_{n}}^{\dagger}}{\Delta \omega} \right] N_{\omega_{n}} \sin \omega_{n} \rho, \qquad \Delta\omega \propto \frac{\pi}{\rho_{s}} \nonumber\\
  &\xrightarrow{\Delta \omega \to 0} \int_{0}^{\infty} d\omega \left[ \mathsf{a}_{\mathrm{r},\omega} + \mathsf{a}_{\mathrm{r},\omega}^{\dagger} \right] N_{\omega} \sin \omega \rho, \nonumber\\
  &\text{where } \mathsf{a}_{\mathrm{r},\omega} \equiv \lim_{\Delta \omega \to 0} \frac{a_{\mathrm{r},\omega_{n}}}{\Delta\omega}.
  \label{eqn:ctm-limit}
\end{align}

The algebra at $\delta = 0$, then, is generated by combinations of $\mathsf{a}_{\mathrm{r},\omega}$.
Of course, the operator $\mathsf{a}_{\mathrm{r},\omega}$ itself is not a well-defined operator, because
\begin{equation}
  \left[ \mathsf{a}_{\mathrm{r},\omega}, \mathsf{a}_{\mathrm{r},\omega'}^{\dagger} \right] = \delta(\omega - \omega') \qquad \implies \qquad || \mathsf{a}_{\mathrm{r},\omega}^{\dagger} \ket{\psi} ||^{2} = \mel{\psi}{\mathsf{a}_{\mathrm{r},\omega} \mathsf{a}_{\mathrm{r},\omega}^{\dagger}}{\psi} \approx \delta(0) \not\in \mathds{R}^{+}.
  \label{eqn:ctm-ladd-bad}
\end{equation}
Thus, these operators are not maps from a Hilbert space to itself.
So, in this limit, the algebra of the right is made out of the creation and annihilation operators integrated against sufficiently smooth functions $f(\omega)$, which is qualitatively different from the algebra at $\delta > 0$.
This algebra, of course, is type $III_{1}$.
Note that the type $I$ algebras at finite $\delta$ are \emph{not} subalgebras of the type $III$ algebra.

\subsection{Generalities} \label{ssec:type-3-lim}
In what follows, I will use two notions relating to a local QFT algebra $\mathcal{a}_{\mathrm{r}}$.
These notions are not precisely defined here.
Since I'll only use them to understand the physics of the result and not for the result itself, the extra technical baggage of precise definitions would not improve this work.

\paragraph{Type $I$ Approximation}
This is an algebra $\mathcal{a}_{\mathrm{r}}^{(\delta)}$, which is type $I$ and has the following properties.
First, the modular Hamiltonian for a particular special state has level spacing $\sim \log \tfrac{1}{\delta}$.
Second, there is a shared notion of locality between $\mathcal{a}_{\mathrm{r}}$ and $\mathcal{a}_{\mathrm{r}}^{(\delta)}$, such that operators in $\mathcal{a}_{\mathrm{r}}$ smeared over a length scale $\gg \delta$ are well-approximated by operators in $\mathcal{a}_{\mathrm{r}}^{(\delta)}$.\footnote{Local algebras in QFT are believed to all be hyper-finite, which means that they can be approximated by type $I$ algebras. These factors are subalgebras however, and that is not what I am demanding.}
I have not defined this second point in a rigorous way.
I am \emph{not} requiring that the type $I$ approximation be a subalgebra --- otherwise, the finite $\delta$ algebras in the example above would not be type $I$ approximations.

Examples of such approximations are the algebras of the subregion after latticisation, or the example given above using a brick wall.
It should be noted here that there is a notion of \emph{optimal type $I$ approximation} introduced by \cite{Jafferis:2019wkd,Hung:2019bnq} (they call it an isometric factorisation map or a shrinkable boundary condition).
This is the type $I$ approximation (also depending on an arbitrary length scale $\delta$) that preserves all expectation values of a subalgebra (also depending on $\delta$) of operators.
In particular, the optimal type $I$ approximation is \emph{not} a type $I$ factor but generically has a centre.
In the scalar example above, one has to integrate over all Dirichlet boundary conditions at the stretched horizon \cite{Hung:2019bnq}.

\paragraph{Type $III$ Limit}
The second notion is an inverse of the first.
Consider a series of type $I$ algebras $\left( \mathcal{a}_{\mathrm{r}}^{(\delta)} \big| 0 < \delta \ll 1 \right)$ such that they are all approximations of the \emph{same} type $III_{1}$ algebra $\mathcal{a}_{\mathrm{r}}$.
Given this set of type $I$ algebras, the type $III$ limit is just $\mathcal{a}_{\mathrm{r}}$.

\section{The Construction}  \label{sec:main-construction}
I now describe the central construction of this work.
This construction involves taking a situation where the local algebra of one side is a type $I$ non-factor algebra and massaging it into a form reminiscent of the crossed-product algebra.
More precisely, I construct a trace on an algebra $\mathcal{a}_{\mathrm{R}}$ that is a direct sum of factors $\oplus_{\alpha} \mathcal{a}_{\mathrm{r},\alpha}$\footnote{The case of a factor follows by only taking one value of $\alpha$. Subtleties will be dealt with in footnotes.} that satisfies the following properties:
\begin{enumerate}
  \item When each factor is type $I$, this trace is a renormalised version of the usual trace.
  \item When the factors are isomorphic and approximately type $III$, this renormalised trace approximates the trace for the type $II$ crossed-product algebra.
\end{enumerate}

\subsection{A General Type $I$ Algebra} \label{ssec:gen-const}
A type $I$ factor is the algebra of bounded operators on a Hilbert space $B(\mathcal{H}_{\mathrm{r}})$.
We are interested in a general type $I$ algebra, which is a direct sum over type $I$ factors
\begin{equation}
  \mathcal{a}_{\mathrm{R}} = \bigoplus_{\alpha} \mathcal{a}_{\mathrm{r},\alpha}, \qquad \mathcal{a}_{\mathrm{r},\alpha} = B \left( \mathcal{H}_{\mathrm{r},\alpha} \right).
  \label{eqn:phys-alg}
\end{equation}
Here, $\alpha$ labels the sector, and this algebra does not contain any operators that change the sector $\alpha$.
There is a non-trivial centre\footnote{
	A centre $\mathcal{Z}$ of an algebra $\mathcal{A}$ is a subalgebra that commutes with every operator in the parent algebra, $[\mathcal{Z}, \mathcal{A}] = 0$.
	A factor is an algebra with a trivial centre (consisting of multiples of the identity).
} consisting of operators of the form $\oplus_{\alpha} f(\alpha) \mathds{1}_{\alpha}$, where $f$ is a complex-valued function of the sector.
This structure occurs in lattice gauge theories \cite{Donnelly:2011hn,Casini:2013rba,Soni:2015yga} as well as holographic QECs \cite{Harlow:2016vwg} (see figure \ref{fig:hol-qec}).

This algebra, like every von Neumann algebra, can be realised as a subalgebra of operators on a larger Hilbert space $B(\mathcal{H})$.
This larger Hilbert space has the form
\begin{equation}
  \mathcal{H} = \bigoplus_{\alpha} \mathcal{H}_{\mathrm{l},\alpha} \otimes \mathcal{H}_{\mathrm{r},\alpha} \subset \mathcal{H}_{\mathrm{ext}} = \bigoplus_{\alpha} \mathcal{H}_{\mathrm{l},\alpha} \otimes \mathcal{H}_{\mathrm{r},\alpha} \otimes \mathcal{H}_{\mathrm{fus},\alpha}
  \label{eqn:ehs}
\end{equation}
where $\mathcal{H}$ is the subspace of $\mathcal{H}_{\mathrm{ext}}$ that looks like
\begin{equation}
  \ket{\Psi} = \sum_{\alpha} \sqrt{p_{_\alpha}} \ket{\psi}_{\mathrm{lr},\alpha} \otimes \ket{\chi_{\alpha}}, \qquad \ket{\psi}_{\mathrm{lr},\alpha} \in \mathcal{H}_{\mathrm{l},\alpha} \otimes \mathcal{H}_{\mathrm{r},\alpha},\ \ket{\chi_{\alpha}} \in \mathcal{H}_{\mathrm{fus},\alpha}.
  \label{eqn:ehs-state}
\end{equation}
Here, $\left\{ \ket{\chi_{\alpha}} \right\}$ is a fixed set of states, one for every $\alpha$.
The states appearing in \eqref{eqn:ehs-state} are normalised as
\begin{equation}
  \braket{\psi}{\psi}_{\alpha} = 1, \qquad \braket{\chi_{\alpha}}{\chi_{\alpha}} = 1.
  \label{eqn:ehs-state-norm}
\end{equation}

The extra Hilbert space $\mathcal{H}_{\mathrm{fus},\alpha}$ in \eqref{eqn:ehs} is known as a `fusion tensor factor,' a name that comes from the lattice gauge theory literature.
It has a tensor factorisation
\begin{equation}
  \mathcal{H}_{\mathrm{fus},\alpha} = \mathcal{H}_{\mathrm{fus},\mathrm{L},\alpha} \otimes \mathcal{H}_{\mathrm{fus},\mathrm{R},\alpha}.
  \label{eqn:fusion-tensor-fact}
\end{equation}
The state $\ket{\chi_{\alpha}}$ is a fixed state.\footnote{Note that it is possible to have this fusion tensor factor even in the case of a factor, i.e. when $\alpha$ takes only one value.}
The fact that $\mathcal{a}_{\mathrm{R}}$ cannot change the value of $\alpha$ (i.e. that it has the direct sum structure in \eqref{eqn:phys-alg}) descends here from the fact that $\alpha$ can be measured also in the left algebra $\mathcal{a}_{\mathrm{L}} = \mathcal{a}_{R}'$.
The choice of $\mathcal{H}_{\mathrm{fus},\alpha}$ and $\ket{\chi_{\alpha}}$ are entirely arbitrary from the algebraic point-of-view, since the algebras of interest $\mathcal{a}_{\mathrm{L},\mathrm{R}}$ do not have access to them.
This arbitrariness arises when we represent these algebras on a Hilbert space; they are not there in the algebras themselves.
We can think of them as representation ambiguities.

Some notation: I will use small letters $\mathrm{l/r}$ for the `bulk' subsystems (the subalgebras acting on each $\mathcal{H}_{\mathrm{l/r},\alpha}$) and capital letters $\mathrm{L,R}$ for the `boundary' subsystem (that includes the central operator).
The nomenclature of `bulk' and `boundary' is motivated by analogy with the holographic QEC \cite{Harlow:2016vwg}; heuristically, the centre operator is available to a boundary observer but not a bulk observer.
Note that there is no bulk algebra $\mathcal{a}_{\mathrm{r}}$ here, only the bulk algebras $\mathcal{a}_{\mathrm{r},\alpha}$ for each sector; in section \ref{ssec:limit} we will make all the $\mathcal{a}_{\mathrm{r},\alpha}$s isomorphic and then there will be a bulk algebra $\mathcal{a}_{\mathrm{r}}$.
I will also put a hat on top of operators when I am using the same letter for the operator and its eigenvalue.

As mentioned above, the choice of $\mathcal{H}_{\mathrm{fus},\alpha}$ and also $\ket{\chi_{\alpha}}$ is not unique given only $\mathcal{a}_{\mathrm{R}}$.
Both in lattice gauge theories as well as in holographic QECs \cite{Akers:2018fow} $\ket{\chi_{\alpha}}$ is maximally entangled across the factorisation in \eqref{eqn:fusion-tensor-fact}, and I will assume this to be the case.\footnote{This is not so in quantum group lattice gauge theories or more generally if the trace with the correct adjoint-invariance properties is a `quantum trace' --- i.e. $\Tr O = \tr \left( \mathds{D} O \right)$, where $\tr$ is the naive trace and $\mathds{D}$ is known as the defect operator or the Drin'feld element depending on context --- a situation that is relevant for gravity \cite{Jafferis:2019wkd,Donnelly:2020teo,Mertens:2022ujr,Wong:2022eiu,Akers:2023}. Repeating the analysis of \cite{Akers:2018fow} with the quantum trace leads to a non-maximally entangled state \cite{Soni:-1qdef}. In this case $S_{\alpha}$ defined in \eqref{eqn:ee-decomp} becomes a non-trivial operator even within each sector . My analysis uses pretty heavily that it is a number; it would be interesting to address this lacuna. \label{fnote:qgp}}
The entanglement entropy does depend on this choice; it is
\begin{equation}
  S_{E} (\Psi, \mathrm{R}) = \sum_{\alpha} p_{\alpha} \left( - \log p_{\alpha} - \tr \rho_{\alpha} \log \rho_{\alpha} + S_{\alpha} \right), \qquad S_{\alpha} \equiv \log \dim \mathcal{H}_{\mathrm{fus},\alpha}.
  \label{eqn:ee-decomp}
\end{equation}
It is crucial here that $\alpha$ is a discrete index.
In the holographic QEC, this is $S_{\alpha} \sim A/4 G_{N}$.

Given that $S_{\alpha}$ is the only effect of these $\ket{\chi_{\alpha}}$s, let me now choose the indices $\alpha$ to be\footnote{$\alpha$ is a label and can be chosen to be any property of the corresponding sector. I'm choosing this property to be $S_{\alpha}$.}
\begin{equation}
  \alpha = S_{\alpha}
  \label{eqn:alpha-name}
\end{equation}
and replace $\ket{\chi_{\alpha}}$ by
\begin{equation}
  \ket{\chi_{\alpha}} \to \ket{\alpha}_{\mathrm{L}} \ket{\alpha}_{\mathrm{R}}.
  \label{eqn:new-state}
\end{equation}
These steps are valid so long as there are no degeneracies in $S_{\alpha}$; I assume this to be the case for simplicity, but nothing important depends on this assumption.
This is analogous to a passage from the CFT Hilbert space to a semi-classical Hilbert space in AdS/CFT; low-energy measurements cannot resolve the $e^{S}$-dimensional microcanonical subspace and can only measure $S$ --- or, more precisely, the closely related quantity of `boundary' energy.
Introduce the modified trace of a right-sided operator $a = \oplus_{\alpha} a_{\alpha}$ \cite{Akers:2023fqr}
\begin{equation}
  \tr a = \sum_{\alpha} e^{\alpha} \tr_{\mathcal{H}_{\mathrm{r},\alpha}} a_{\alpha}.
  \label{eqn:trace}
\end{equation}
Define also the `area' operators
\begin{equation}
  \hat{\alpha}_{\mathrm{L/R}} = \alpha \ket{\alpha}_{\mathrm{L/R}} \bra{\alpha}.
  \label{eqn:alpha-op}
\end{equation}

Now fix a base state\footnote{
	This is normalised as written if we modify the trace as in \eqref{eqn:trace} \emph{without} modifying the inner product, so that $\braket{\alpha}{\alpha} = 1$ but $\tr \ket{\alpha} \bra{\alpha} = e^{\alpha}$.
	This is a slightly unnatural, but more convenient, choice.
}
\begin{equation}
  \ket{\Phi} = \sum_{\alpha} \sqrt{q_{\alpha}} \ket{\varphi}_{\mathrm{lr},\alpha} \ket{\alpha}_{\mathrm{L}} \ket{\alpha}_{\mathrm{R}}.
  \label{eqn:base-state}
\end{equation}
This is the state I will use to rewrite the trace \eqref{eqn:trace} using \eqref{eqn:tr-triv}.
It will be useful to use the Schmidt basis for $\ket{\varphi}_{\alpha}$
\begin{equation}
  \ket{\varphi}_{\alpha} = \sum_{k_{\alpha}} e^{- \frac{k_{\alpha}}{2}} \ket{k_{\alpha}}_{\mathrm{l}} \ket{k_{\alpha}}_{\mathrm{r}}, \qquad \hat{k}_{\alpha,\mathrm{l/r}} = \sum_{k_{\alpha}} k_{\alpha} \ket{k_{\alpha}}_{\mathrm{l/r}} \bra{k_{\alpha}}.
  \label{eqn:sector-schmidt}
\end{equation}
$\ket{\Phi}$ is chosen so that these are complete bases of $\mathcal{H}_{\mathrm{l/r},\alpha}$ and none of the $k_{\alpha}$s vanish.
I have also defined the one-sided modular Hamiltonians $\hat{k}_{\alpha,\mathrm{l/r}}$ here. 
The one-sided modular Hamiltonian for $\Phi$ is
\begin{equation}
  \hat{K}_{\Phi,\mathrm{R}} = \sum_{\alpha} \left( \alpha - \log q_{\alpha} + \hat{k}_{\alpha,\mathrm{r}} \right) \ket{\alpha}_{\mathrm{r}} \bra{\alpha}.
  \label{eqn:full-K-base}
\end{equation}
Applying \eqref{eqn:tr-triv} at this stage will not result in anything insightful.

The fundamental reason for this is that at this stage the algebra is made out of a `bulk' operator for every value of the `area' whereas the crossed product algebra is one `bulk' operator for every value of the left `boundary' Hamiltonian.
For $\alpha \in \spec \hat{\alpha}_{\mathrm{L}}, k_{\alpha} \in \spec \hat{k}_{\alpha,\mathrm{l}}$, define the `boundary Hamiltonian' $X$, the probability $q_{X}$ of measuring the value $X$ and the set $\mathsf{K} (X)$ as the simultaneous solutions of the following equations
\begin{align}
  X - \log q_{X} &= \alpha + k_{\alpha} - \log q_{\alpha} \nonumber\\
  k_{X} \in \mathsf{K}(X) \quad \Leftrightarrow \quad q_{X} e^{- k_{X}} &= q_{\alpha} e^{- k_{\alpha}} \nonumber\\
  \sum_{k_{X} \in \mathsf{K}(X)} e^{- k_{X}} &= 1.
  \label{eqn:X-defn}
\end{align}
These equations are the discrete versions of the JLMS relation \cite{Wald:1993nt,Jafferis:2015del}; they will become more familiar in section \ref{ssec:limit}.
We remind the reader that, despite the name, $X$ is not a Hamiltonian governing the real-time evolution of any system but the modular Hamiltonian in a putative boundary dual (more precisely, it has the same eigenvalues as the modular Hamiltonian but is not the same operator).

Call the set of solutions for $X$ as $\Xi$.\footnote{In the case of a factor, $q_{\alpha} = 1$ and $\Xi = \left\{ \alpha + k \big| k \in \spec \hat{k} \right\}$. Assuming no degeneracies in the entanglement spectrum of $\ket{\Phi}$ --- a choice that can always be made --- $q_{X} = e^{- X + \alpha}, k_{X} = 0$.}
With these definitions, the basis elements of the Hilbert space can be rewritten as
\begin{equation}
  \ket{k_{\alpha}}_{\mathrm{l}} \ket{\alpha}_{\mathrm{L}} = \ket{X}_{\mathrm{L}} \ket{k_{X}}_{\mathrm{l}}, \qquad \ket{\alpha}_{\mathrm{R}} \ket{k_{\alpha}}_{\mathrm{r}} = \ket{k_{X}}_{\mathrm{r}} \ket{X}_{\mathrm{R}}.
  \label{eqn:basis-relabel}
\end{equation}
In terms of this new basis, the base state is
\begin{align}
  \ket{\Phi} &= \sum_{\alpha,k_{\alpha}} \sqrt{q_{\alpha}} e^{- \frac{k_{\alpha}}{2}} \ket{k_{\alpha}}_{\mathrm{l}} \ket{\alpha}_{\mathrm{L}} \ket{\alpha}_{\mathrm{R}} \ket{k_{\alpha}}_{\mathrm{r}} \nonumber\\
  &= \sum_{X \in \Xi} \sum_{k_{X} \in \mathsf{K}(X)} \sqrt{q_{X}} e^{- \frac{k_{X}}{2}} \ket{X}_{\mathrm{L}} \ket{k_{X}}_{\mathrm{l}} \ket{k_{X}}_{\mathrm{r}} \ket{X}_{\mathrm{R}} \nonumber\\
  &\equiv \sum_{X \in \Xi} \sqrt{q_{X}} \ket{X}_{\mathrm{L}} \ket{\varphi}_{\mathrm{lr},X} \ket{X}_{\mathrm{R}}.
  \label{eqn:base-state-final}
\end{align}
This also clarifies that the set $\mathsf{K}(X) = \spec \hat{k}_{X,\mathrm{l}}$, where the latter is the one-sided modular Hamiltonian for $\ket{\varphi}_{X}$.

A general state in the full Hilbert space gets renamed to
\begin{equation}
  \ket{k_{\alpha}}_{\mathrm{l}} \ket{\alpha}_{\mathrm{L}} \ket{\alpha}_{\mathrm{R}} \ket{k'_{\alpha}}_{\mathrm{R}} = \ket{X}_{\mathrm{L}} \ket{k_{X}}_{\mathrm{l}} \ket{k_{X'}}_{\mathrm{r}} \ket{X'}_{\mathrm{R}}.
  \label{eqn:stuff}
\end{equation}
There is a relation between $X$ and $X'$ due to the equality of $\alpha$s on the left and the right.
Notice first from \eqref{eqn:X-defn} that
\begin{equation}
  \alpha = X + \log \left[ \frac{q_{\alpha}}{q_{X}} e^{- k_{\alpha}} \right] = X - k_{X}.
  \label{eqn:area-adm-reln}
\end{equation}
As a result
\begin{equation}
  X' - X = k_{X'} - k_{X}.
  \label{eqn:gauge-cons}
\end{equation}
Abstractly,
\begin{equation}
  \bigoplus_{\alpha} \mathcal{H}_{\alpha,\mathrm{l}} \otimes \mathcal{H}_{\alpha,\mathrm{r}} = \left. \bigoplus_{X,X' \in \Xi} \mathcal{H}_{X,\mathrm{l}} \otimes \mathcal{H}_{X',\mathrm{r}} \right|_{X' - X = \hat{k}_{X',\mathrm{r}} - \hat{k}_{X,\mathrm{l}}},
  \label{eqn:full-H-relabel}
\end{equation}
where the operator equation in the subscript is a constraint satisfied by all states.

After similar manipulations, the one-sided modular Hamiltonian \eqref{eqn:full-K-base} for $\Phi$ can be written as
\begin{equation}
  \hat{K}_{\Phi,\mathrm{R}} = \hat{X}_{\mathrm{R}} - \log q_{\hat{X}_{\mathrm{R}}},
  \label{eqn:full-K-base-X}
\end{equation}
where $\hat{X}_{\mathrm{R}},q_{\hat{X}_{\mathrm{R}}}$ measure the value of $X',q_{X'}$ in \eqref{eqn:full-H-relabel}.
The trace \eqref{eqn:tr-triv} is then
\begin{equation}
  \tr a = \mel**{\Phi}{a\, \frac{e^{\hat{X}_{\mathrm{R}}}}{q_{\hat{X}_{\mathrm{R}}}}}{\Phi} = \mel**{\Phi}{a\, \frac{e^{\hat{X}_{\mathrm{L}}}}{q_{\hat{X}_{\mathrm{L}}}}}{\Phi}.
  \label{eqn:tr-2}
\end{equation}
The second equation follows from the constraint \eqref{eqn:gauge-cons} and the fact that $\ket{\varphi}_{X}$ is annihilated by its two-sided modular Hamiltonian $\hat{k}_{X}$.
One can also renormalise it
\begin{equation}
  \Tr a = e^{- X_{0}} \tr a
  \label{eqn:tr-2-ren}
\end{equation}
for some $X_{0}$.
This is remarkably similar to the trace written down in the crossed product algebra; in particular, it neither uses a sum over states in a local tensor factor nor the modular Hamiltonian of $\ket{\varphi}_{\mathrm{lr}}$.
This trace continues to make sense after replacing the local subalgebra at fixed $X$, $B (\mathcal{H}_{X,\mathrm{r}})$, with a type $III_{1}$ algebra, meaning that the local subalgebra including the $X$ degree of freedom would be a type $II$ algebra.

An important difference from the crossed product, however, is that the right subalgebra $\mathcal{a}_{\mathrm{R}}$ from \eqref{eqn:phys-alg} is not block diagonal in $X$,
\begin{align}
  \oplus_{\alpha} a_{\alpha} &= \sum_{\alpha,k_{\alpha},k_{\alpha}'} a_{\alpha,k_{\alpha},k'_{\alpha}} \ket{\alpha,k_{\alpha}} \bra{\alpha,k_{\alpha}'} \nonumber\\
  &= \sum_{X,X',k_{X},k_{X'}} \delta_{X', X + k_{X'} - k_{X}}\ a_{X,k_{X};X',k_{X'}} \ket{X,k_{X}} \bra{X',k_{X'}}.
  \label{eqn:discrete-problem}
\end{align}
The Kronecker delta here can be derived from the same steps used to derive \eqref{eqn:gauge-cons}.
In particular, this is also true of density matrices of states $\ket{\Psi} \neq \ket{\Phi}$.

The problem is that in this equation $X$ is the right `boundary' Hamiltonian, which is not central since everything is dressed to the right `boundary.'
Removing the explicit $\ket{X}_{\mathrm{R}}$ factor from the Hilbert space and using the constraint \eqref{eqn:gauge-cons} to define $\hat{X}_{\mathrm{R}} = \hat{X}_{L} + \hat{k}_{\hat{X}_{\mathrm{R}},\mathrm{r}} - \hat{k}_{\hat{X}_{\mathrm{L}},\mathrm{l}}$, the operator is block-diagonal in $X_{L}$ space.
Before being more explicit about this, it will be useful to specialise to a situation closer to holography.

One has to be more careful for operators on the left; the point is that when we change the dressing as in figure \ref{fig:re-dressing}, the operators on the right stay the same but those on the left \emph{change} --- a left operator dressed to the right boundary is not block diagonal with respect to $\alpha$.
But to build the crossed product algebra for the left the most convenient choice is to dress the operators to the left boundary \cite{Witten:2021unn,Chandrasekaran:2022eqq}.
This means that the actual algebras used are the ones on the left of figure \ref{fig:re-dressing} --- I index the right (left) operator by the left (right) `boundary' energy $X_{\mathrm{L}}$ ($X_{\mathrm{R}}$).
This is exactly the original type $I$ algebra on each side, as discussed already in section \ref{sec:idea}, providing some moral support for the idea that the crossed product is a formal way of dealing with centres in the type $III$ case.

\subsection{The Type $III$ Limit} \label{ssec:limit}
A closer analogy to the crossed product construction is easier to describe in a special `semi-classical' situation.
This situation is one where the following properties are true:
\begin{enumerate}
  \item There is a large parameter $S$. In gravity, $S \sim \frac{1}{G_{N}}$.
  \item $\spec \hat{\alpha}$ is almost continuous, with a spacing $e^{-S}$.
  \item $q_{\alpha}$ is peaked around $\alpha = \alpha_{0} = \mathcal{O} (S)$, with $\mathcal{O} \left( 1 \right)$ width. In other words, I work with a microcanonical ensemble.

    Further, the width of the probability distribution in all states of interest is $\mathcal{O} \left( 1/\varepsilon \right)$, where $\varepsilon = \mathcal{O} \left( S^{0} \right)$ is a different small parameter \cite{Chandrasekaran:2022eqq}.
  \item The Hilbert spaces $\mathcal{H}_{\mathrm{l/r},\alpha} = \mathcal{H}_{\mathrm{l/r}}$ are the same for each $\alpha$, and infinite-dimensional. And therefore the same is true of the algebras $\mathcal{a}_{\mathrm{l/r},\alpha} = \mathcal{a}_{\mathrm{l/r}}$.

    This allows me to choose also a base state $\ket{\Phi}$ such that the `bulk' state $\ket{\varphi_{\alpha}} = \ket{\varphi}$ is also independent of $\alpha$.
  \item $\mathcal{a}_{\mathrm{l/r}}$ is a type $I$ approximation in the sense of section \ref{sec:type-1} with small parameter $\delta \sim \mathcal{O} \left( S^{0} \right), \delta \ll \varepsilon$.
    In particular, this means that the spectrum of the one-sided modular Hamiltonian $\hat{k}_{\mathrm{l/r}}$ of the base state in each sector $\ket{\varphi}$ ranges from $0$ to $\infty$ with spacing $\log \frac{1}{\delta}$; this is true of any state.
    The spectrum of the full two-sided modular Hamiltonian $\hat{k}$ is a discretisation of the full real line.
    This is also true for every other state.\footnote{Note here that I'm not using the non-rigorous second requirement of the type $I$ approximation in section \ref{ssec:type-3-lim}.}

    I will assume that every number in $\spec \hat{k}$ is $\mathcal{O} \left( S^{0} \right)$ even though the spectrum is unbounded above and below; cutting off such large eigenvalues only changes the state by an exponentially small amount.
\end{enumerate}

I now treat $\alpha$ as a continuous variable, and take $\ket{\alpha}$ to be $\delta$-function normalised.
As long as $\alpha$ is in reality discrete, the reader can read $\delta(0) ``=" \mathcal{O} \left( e^{S} \right)$.
With these approximations, the equations \eqref{eqn:X-defn} simplify substantially.
The first equation is
\begin{align}
  X - \alpha &= k + \log \frac{q(X)}{q(\alpha)} = \mathcal{O} (1) \nonumber\\
  \implies \qquad q(\alpha) &= q(X) + (\alpha - X) q' (X) + \cdots = q(X) \left( 1 + \mathcal{O} \left( \varepsilon \right) \right) \nonumber\\
  \implies \qquad X - \alpha &= k + \mathcal{O} \left( \varepsilon \right), \nonumber\\
  \text{and} \qquad k_{\alpha} &\approx k_{X}.
  \label{eqn:X-defn-simple}
\end{align}
The intricate recursive structure of \eqref{eqn:X-defn} has disappeared; it should be noted that the assumptions (apart from the introduction of $\varepsilon$) were chosen to mock up the semi-classical limit of a holographic QEC, \emph{not} to make this simplification happen.

\eqref{eqn:X-defn-simple} implies that $X$ is also approximately continuous.
Thus, the range of values of $\alpha$ and $X$ become the same and the relabelling can be thought of also as a shift
\begin{equation}
  \ket{k}_{\mathrm{l}} \ket{\alpha}_{\mathrm{L}} \ket{\alpha}_{\mathrm{R}} \ket{k'}_{\mathrm{r}} = U_{\mathrm{dr}}^{\dagger} e^{i \hat{k} \hat{\Pi}_{\mathrm{R}}} \ket{k}_{\mathrm{l}} \ket{X}_{\mathrm{L}} \ket{X}_{\mathrm{R}} \ket{k'}_{\mathrm{r}}
  \label{eqn:area-to-adm-ctm}
\end{equation}
where $\hat{\Pi}$ is the conjugate of $\hat{X}$ and the re-dressing unitary $U_{\mathrm{dr}}$ is
\begin{equation}
  U_{\mathrm{dr}} = e^{i \left( \hat{k}_{\mathrm{l}} \hat{\Pi}_{\mathrm{L}} + \hat{k}_{\mathrm{r}} \hat{\Pi}_{\mathrm{R}} \right)},
  \label{eqn:U-cr-pr}
\end{equation}
This re-dressing unitary exactly takes the value of the area to the value of the `boundary' Hamiltonian by adding the modular energy of the bulk fields.
I will call the RHS the left-boundary-indexed state and the LHS the HRT-indexed state.

The $e^{i \hat{k} \hat{\Pi}_{\mathrm{R}}}$ on the RHS imposes the constraint \eqref{eqn:gauge-cons}.
In the limit where $\spec \hat{X}_{\mathrm{L}}$ becomes genuinely continuous, the state on the RHS has norm $\delta(0)$.
This can be dealt with either by taking the dual Hilbert space to be a Hilbert space of co-invariants \cite{Marolf:2008hg,Chandrasekaran:2022cip} or by removing the $\ket{X}_{\mathrm{R}}$ factor and \emph{defining}
\begin{equation}
  \hat{X}_{\mathrm{R}} = \hat{X}_{L} + \hat{k}.
  \label{eqn:gauge-cons-soln}
\end{equation}
This latter is closer to the usual crossed product construction, and this is the one that is more convenient here.
With this change,
\begin{equation}
  U_{\mathrm{dr},\mathrm{R}} = e^{i \hat{k}_{\mathrm{l}} \hat{\Pi}}, \qquad \ket{\alpha} \ket{k}_{\mathrm{l}} \ket{k'}_{\mathrm{r}} = U_{\mathrm{dr,R}}^{\dagger} \ket{X} \ket{k}_{\mathrm{l}} \ket{k'}_{\mathrm{r}}.
  \label{eqn:ctm-limit-2}
\end{equation}
For left-sided operators, we need to dress to the left boundary using
\begin{equation}
  U_{\mathrm{dr,L}} = e^{i \hat{k}_{\mathrm{r}} \hat{\Pi}}.
  \label{eqn:left-bd-dr}
\end{equation}
The confusing aspect of this notation is that in different equations the same tensor factor carries the value of the area or the left or the right `boundary' Hamiltonians.

Since the algebra in the HRT-indexing is made out of operators of the form $\oplus_{\alpha} a_{\alpha}$, that in the boundary-indexing is made out of the same operators conjugated by the re-dressing unitary,
\begin{align}
  U_{\mathrm{dr,R}} \left[ \sum_{\alpha,k,k'} \ket{\alpha,k} a_{\alpha,k,k'} \bra{\alpha,k'} \right] U_{\mathrm{dr,R}}^{\dagger} = \sum_{X} \ket{X,k} a_{X,k,k'} \bra{X, k'}
  \label{eqn:op-redress}
\end{align}
An analogous operation is possible in the general case, but harder to describe, see the discussion at the end of section \ref{ssec:gen-const}.

Another issue that needs to be dealt with is that the trace \eqref{eqn:type-2-tr} diverges in the $S \to \infty$ limit, because there is an $e^{X}$ in it and $X = \mathcal{O} (S)$.
This is easily dealt with by renormalising it.
Define $\delta X = X - X_{0}$; the renormalised trace is then
\begin{equation}
  \Tr a \equiv e^{- X_{0}} \tr a = \mel**{\Psi}{a\, \frac{e^{\hat{\delta X}}}{q (\hat{X})}}{\Psi}.
  \label{eqn:ren-tr}
\end{equation}
This is in line with the idea that the trace in the type $II$ algebra is a renormalised trace.

Now, to show that there is a canonical isomorphism between the crossed product for different base states, simply note that the re-dressing unitary is
\begin{equation}
  U_{\mathrm{dr,R}} = \left( \rho_{\varphi,\mathrm{l}} \right)^{- i \hat{\Pi}}.
  \label{eqn:U-dr-2}
\end{equation}
Therefore the re-dressing unitary based on a different state $\ket{\Psi}$ is
\begin{equation}
  U'_{\mathrm{dr,R}} = \left( \rho_{\psi,\mathrm{l}} \right)^{- i \hat{\Pi}} = \left[ \left( \rho_{\psi,\mathrm{l}} \right)^{- i \hat{\Pi}}\left( \rho_{\varphi,\mathrm{l}} \right)^{i \hat{\Pi}} \right] U_{\mathrm{dr,R}}.
  \label{eqn:U-dr-3}
\end{equation}
The operator in the square brackets is exactly the Connes cocycle unitary used for this isomorphism in \cite{Witten:2021unn}.
In the general case, this independence is clear simply because all I did there was rename states; recasting that renaming as the action of a unitary here makes it seem non-trivial.

This isomorphism was also used in \cite{Chandrasekaran:2022eqq} to show that the type $II$ entropy was the difference of generalised entropy from the base state.
This proof applies to this situation as well, since it differs from the continuum due to the finiteness of $S,\delta^{-1} \gg \varepsilon^{-1}$ and the proof of \cite{Chandrasekaran:2022eqq} has errors of $\mathcal{O} \left( \varepsilon \right)$.

Finally, the conditional expectation of \cite{Gesteau:2023hbq} exactly leaves the total modular energies $X,X'$ (more precisely, $\spec \hat{X}_{\mathrm{L/R}}$) unchanged while changing $\spec \hat{\alpha}$ and $\spec \hat{k}_{\mathrm{l/r}}$.
The trace, then, is manifestly invariant under this operation.
This conditional expectation can be thought of as coarse-graining the type $I$ approximation (e.g. $\delta \to 2\delta$ in the example of section \ref{sec:type-1}).

\subsection{Type $II_{1}$ Algebras} \label{ssec:2-1}
If the area has a maximum value $\alpha \le \alpha_{0}$, then the `boundary' energy has a maximum value which is $\alpha_{0} + \mathcal{O} \left( S^{0} \right)$.
Taking $X_{0}$ to be the maximum value of $X$ results in $\Tr \mathds{1} = 1$ and therefore the algebra approximates a type $II_{1}$ algebra.

Inverting the logic, along with the fact that the gravitational crossed product is relevant for de Sitter and also general gravitating subregions, indicates that the holographic QEC, or at least the boxed part of figure \ref{fig:hol-qec}, might apply to these cases also.
This corroborates intuition that the area operator is central in general situations; however, if the holographic QEC is relevant in these situations, it is unclear what Hilbert space the unitaries (which are the holographic as well as the error-correcting part of the circuit) map the semi-classical state into.\footnote{I thank Matthew Heydeman for emphasising this last subtlety to me.}

\section{Discussion} \label{sec:conc}
I have shown that the crossed product construction can be approximated in a type $I$ setting.
The main physical insights used can be found in \cite{Chandrasekaran:2022eqq}.
The solution of this `engineering' problem can be used in the context of tensor networks and holographic QECs, but also in the more general context of local algebras with centres like lattice gauge theories.

The most interesting outcomes of the main analysis are:
\begin{enumerate}
  \item The transformation from the holographic QEC to the approximate crossed-product can be thought of as a change of dressing. It would be interesting to make this precise in the language of quantum reference frames.
    It would further be interesting to see if the whole concept of the gravitational crossed product can be usefully recast in that language.
  \item The discussion of section \ref{ssec:2-1} indicates that we should in fact think of the area operator of the boundary of an arbitrary subregion, e.g. the static patch of de Sitter, as a central operator in the regulated semi-classical algebra.
    This again is not a surprise; the interesting statement is that this is justified as a consequence of the success of the gravitational crossed product.
  \item The holographic QEC can contain `non-classical' states, those where the area has $\mathcal{O} \left( 1/G_{N} \right)$ fluctuations. The steps above then lead to a direct sum of approximate crossed products, one for each classical background.
\end{enumerate}

There are some interesting future directions.
\begin{enumerate}
  \item Make the discussion of section \ref{ssec:type-3-lim} more precise.

  This might also help make my construction more elegant.
  \item Extend this construction to the case when the fusion state is not maximally entangled.
    This is relevant for the reasons discussed in footnote \ref{fnote:qgp}.
	\item It would be interesting to make the connection with the description in terms of geometrical phases over the space of states in \cite{Banerjee:2023eew}.
  \item This construction shows that there is a similarity between the crossed product algebra and the code algebra in a holographic QEC.
    The structure of the holographic QEC was found by demanding the Ryu-Takayanagi formula and entanglement wedge reconstruction in type $I$ algebras \cite{Harlow:2016vwg}.
    This suggests the following very interesting question: can the crossed product algebra be related in a similarly precise way to the generalised entropy formula and entanglement wedge reconstruction in the case where the bulk algebra is type $III$, prehaps in an asymptotically isometric setting \cite{Faulkner:2022ada}?
  \item Use this construction to understand the relation between the path integral and algebraic proofs of the finiteness of generalised entropy.\footnote{I thank Shiraz Minwalla for emphasising the importance of this question. I apologise for not having solved it.}
    There are two path integral proofs, both of which assert that the renormalisation of Newton's constant required to make \emph{any} Euclidean path integral finite \emph{also} renders the generalised entropy finite.
    The two proofs differ in the replication scheme.
    The much better studied one, following in the footsteps of \cite{Susskind:1994sm}, is one where a conical singularity is introduced at the horizon.
    Another one \cite{Larsen:1995ax} uses the Gibbons-Hawking path integral \cite{Gibbons:1976ue} and so does not introduce a conical singularity.
    In this second scheme the statement of finiteness is straightforward\footnote{There are subtleties regarding phase transitions and mass thresholds in the bulk QFT, but I believe the same subtleties will exist in the algebraic proof.} but the interpretation of the matter correction as a matter entanglement entropy is \emph{not}.
    Note that the first prescription is entirely wrong if one is interested in R\'enyi entropies, and so it behooves us to study the Gibbons-Hawking prescription.

    This is good news, since as we have seen the crossed product allows the area to fluctuate, a feature shared only by the Gibbons-Hawking prescription does.
    In particular, the calculation of R\'enyi entropy in the crossed product algebra reproduces the answer from the Gibbons-Hawking path integral \cite{Witten:2021unn}.\footnote{A little more work is required to show this including matter corrections. \cite{Witten:2021unn} showed that the crossed product reproduces log corrections to the entropy one expects from the boundary canonical ensemble. These log corrections vanish if one does not allow the area and the ADM energy to fluctuate as in the conical singularity prescription.}
    The holographic QEC is an even closer analog, since it is (a) regulated and (b) natural from the bulk point of view to work in sectors of fixed area rather than fixed ADM energy.
    Further, the same subtlety of identifying the matter contribution as the entanglement entropy of a specific state in the bulk QFT exists --- since changing $X$ also changes the geometry,\footnote{There is also a subleading issue where a single $X$ sector is not exactly a single geometry.} which means that the different states $\ket{\psi}_{X}$ live in different Hilbert spaces.
    In the Gibbons-Hawking path integral, the bulk remains smooth upon replication because the dominant ADM energy changes when we replicate, meaning that it is not just the bulk state on the original geometry that contributes to the matter part.
    \cite{Chandrasekaran:2022eqq} (and, in their footsteps, I) dealt with this subtlety by fixing the ensemble to be microcanonical.
    Since an $\mathcal{O} (1)$ change in the ADM energy is an $\mathcal{O}(G_{N})$ change in the bulk geometry, one does not have to worry about the above subtlety.

    So, to make the connection precise, we would have to either prove the equality of type $II$ and generalised entropies in the canonical ensemble or repeat the analysis of \cite{Larsen:1995ax} in the microcanonical ensemble.
  \item Work out an example in JT gravity.
    This would involve constructing a holographic QEC using a smeared area operator like $\int_{\mathcal{M}_{2}} \phi f_{\delta} (\theta_{+}) f_{\delta} (\theta_{-})$, where $f_{\delta}(x)$ is sharply peaked at $x=0$ and width $\mathcal{O} (\delta)$.
    This is the same $\delta$ as in section \ref{ssec:eg}.
    The main technical obstacle is to find a function $f_{\delta}$ such that the relation \eqref{eqn:X-defn-simple} between ADM energy and smeared area holds.
\end{enumerate}

\section*{Acknowledgements}
I thank participants in the type $II$ algebra study week at DAMTP, University of Cambridge, --- Amr Ahmadain, Goncalo Araujo-Regado, Alex Frenkel, Rifath Khan, Prahar Mitra, Krishnendu Ray, Ayngaran Thavanesan, Bilyana Tomova and Manus Visser ---, especially those who sat through my explanation of this work at a very early stage.
I thank Wissam Chemissany, Matthew Heydeman, David K. Kolchmeyer, Shiraz Minwalla, Onkar Parrikar, Antony Speranza, Gabriel Wong and the audience of a talk at Tata Institute of Fundamental Research (TIFR) for discussions.
I thank Chris Akers and Annie Wei for collaboration on a related project \cite{Akers:2023} that led to this work and comments on previus versions; and also David Tong for asking about the possibility of approximating the crossed product.
Finally, I thank an anonymous reviewer of my talk proposal to the ``It from Qubit 2023'' conference for pointing out an important mistake in my presentation.

This work has been partially supported by STFC consolidated grant ST/T000694/1.
I am supported by the Isaac Newton Trust grant ``Quantum Cosmology and Emergent Time'' and the (United States) Air Force Office of Scientific Research (AFOSR) grant ``Tensor Networks and Holographic Spacetime''.

\bibliographystyle{JHEP}
\bibliography{refs}
\end{document}